# Large scale deployment of C-ITS: Impact assessment results of the C-Roads Greece pilots

**Areti, Kotsi[1], Dr. Evangelos, Mitsakis[1]**

[1]Centre for Research and Technology Hellas (CERTH) - Hellenic Institute of Transport (HIT)
akotsi@certh.gr, emit@certh.gr

**Abstract**

This paper aims to provide insights related to the impact assessment and evaluation results from the use of C-ITS services in the Greek pilot of the C-Roads Greece project, i.e., Attica Tollway and Egnatia Odos Tollway. The impact assessment and evaluation of the C-ITS services includes aspects related to user acceptance, real-world pilot logs collected from the two pilots, and simulation experiments that were conducted for the impact assessment of the C-ITS services. The paper concludes with a roadmap and guidelines for the extended deployment of C-ITS services in the country's highway and urban road networks.

## 1. Introduction

Development and deployment of Cooperative Intelligent Transport Systems (C-ITS) take place with the aim of improving safety, traffic efficiency, energy efficiency and comfort. C-ITS are based on ICT (Information and Communication Technologies), such as sensor technology, telecommunications, information processing and control technology. Various technologies can be combined in different ways to create stand-alone in-vehicle systems and cooperative systems (V2X) (Turetken et al., 2018). The C-Roads platform is an umbrella initiative which aims to harmonize the efforts countries in Europe make to deploy C-ITS. With 16 member states participating as core members, the C-Roads platform pursues cooperation on all levels of C-ITS deployment. This allows sharing of experience and knowledge regarding the implementation and deployment issues (Froetscher & Monschiebl, 2018).

Greece, under the frame of the C-Roads Platform initiative, participates as a member state with its national pilot since June 2019. The Greek pilot takes place at two test sites, one in Northern Greece (Egnatia Odos Tollway Pilot) and one in Attica Region (Attica Tollway Pilot), and its main objective is to deploy a specific set of Day 1 and Day 1.5 C-ITS services by using a balanced mixture of ETSI ITS G5 and cellular communication technologies. A breakdown of the selected Day 1 C-ITS services of the Greek pilot includes: Road Works Warning (RWW), with special focus on lane closure and other restrictions, Hazardous Locations Notification (HLN), which includes the cases of a stationary vehicle, weather condition warning and obstacle on the road, In-Vehicle Signage (IVS), where Variables Message Signs of "free text" as well as speed advice messages to avoid the shockwave damping effect are sent to vehicles by the road operators, Probe Vehicle Data (PVD) in regards to CAM Aggregation, and Smart Routing, which is a Day 1.5 C-ITS service (C-Roads Platform, 2023).

In the context of the C-Roads Platform activities, and more specifically Working Group (WG3): Evaluation and Assessment, the project conducted an impact assessment and evaluation of the C-ITS services that were demonstrated in the road networks of the pilot locations. The impact assessment and evaluation activities of C-Roads Greece were performed in accordance with (Studer et al., 2020), an impact assessment and evaluation plan which purpose is to create the common basis for evaluation and assessment of the C- Roads Pilots.

With regards to C-ITS impact assessment and evaluation activities in real-life environments, Froetscher and Monschiebl (Froetscher & Monschiebl, 2018) describe the different steps undertaken in the living lab environment in Vienna to ensure interoperability and system functionality of C-ITS services in a wide network of independent stakeholders as participants. Lokaj et al. (Lokaj et al., 2020) provide general information about a model of the system developed for C-Roads Czech Republic and its parts, as well as methods for testing and technical evaluation primary based on the log-analysis of the C-ITS communication. Agriesti et al. (Agriesti et al., 2018) report an extended summary of the available bibliography on Truck Platooning and describe an evaluation methodology aiming to assess the jointed impact of Day 1 C-ITS services and Truck Platooning. Studer et al. (Studer et al., 2019) present the approach for impact assessment and evaluation defined by the C-Roads Platform and applied to the Italian pilot. Otto et al. (Otto et al., 2023) propose a framework to evaluate the effectivity of GLOSA (measured as percent of the cycle time where it is available with high confidence), considering the current signal program parameters like cycle time or proportion of green per cycle, traffic flow parameters like traffic load or queue length, and limitations on the lowest and highest possible speed advised by GLOSA, as well as the range of communication depending on the communication technology.

Concerning previous works in the domain of C-ITS services impact assessment and evaluation, Olia et al. (Olia et al., 2016) use a modeling framework to simulate in microscopic scale the traffic of a real network in the city of Toronto, Canada. Park and Lee (Lee & Park, 2008) conduct microsimulation experiments using the VISSIM software, to assess various route guidance strategies. Factors taken into consideration for impacts assessment include market penetration of vehicles with access to C-ITS, congestion levels, updating intervals of route guidance information, and drivers' compliance rates. Dion et al. (Dion & Robinson, 2009) develop a framework for the assessment of V2I-based communications using the Paramics microscopic traffic simulator. More specifically, the authors implement a virtual testbed where they examine the dissemination of messages between vehicles and roadside equipment and the respective generation of probe vehicle data. Kattan et al. (Kattan et al., 2012) focus on the assessment of V2V-based communications with a specific interest in hazardous situations, i.e., incidents resulting from adverse weather conditions. To simulate and test the messages exchange, two Paramics Applications Programming Interfaces (APIs) are used: one to simulate incidents creation and the other one to simulate the message exchange between individual C-ITS enabled vehicles. The test network where various scenarios with moderate and high congestion are tested includes urban arterials. Experiments results indicate the positive impact of V2V technologies in terms of safety and travel time improvement.

## 2. Greek pilot

The C-Roads Greece pilot includes two test sites, one in Northern Greece, the Egnatia Odos Tollway pilot, and one in Attica Region, the Attica Tollways pilot.

## 2.1 Attica Tollways

Attiki Odos Motorway, i.e., Attica Tollway, in Athens, Greece, is a 70 km-long urban motorway, fully access-controlled through 39 toll barriers. Attica Tollway is the Athens Ring Road, providing free-flow traffic conditions in the city centre-periphery, linking the downtown areas by radial connections and main arterials. It provides a link between the national motorway network to the south and the north of the Greek Capital and connects the city and its suburbs with the new Athens International Airport. The Attica Tollway pilot involves approximately 21 km of the central sector of Attica Tollway (motorway consisting of the ring road of the greater metropolitan area of Athens) along the main axis of the road, between the Dimokratias Interchange and Paiania Interchange (from km 21 to km 42 of the Attica Tollway), corresponding to the section of the motorway with the heaviest traffic (53,000 vehicles as per data registered in 2018). The already installed ITS related equipment (Variable Message Signs, CCTV traffic cameras, traffic detection inductive loops, meteo and smoke sensors) was utilized alongside the C-ITS field equipment that was installed during the pilot.

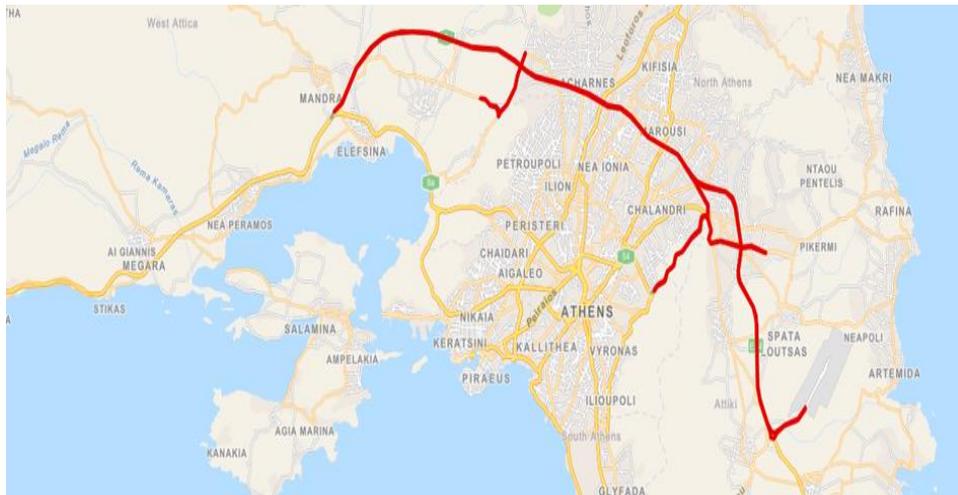

*Figure 1: Map of the Attica Tollway pilot*

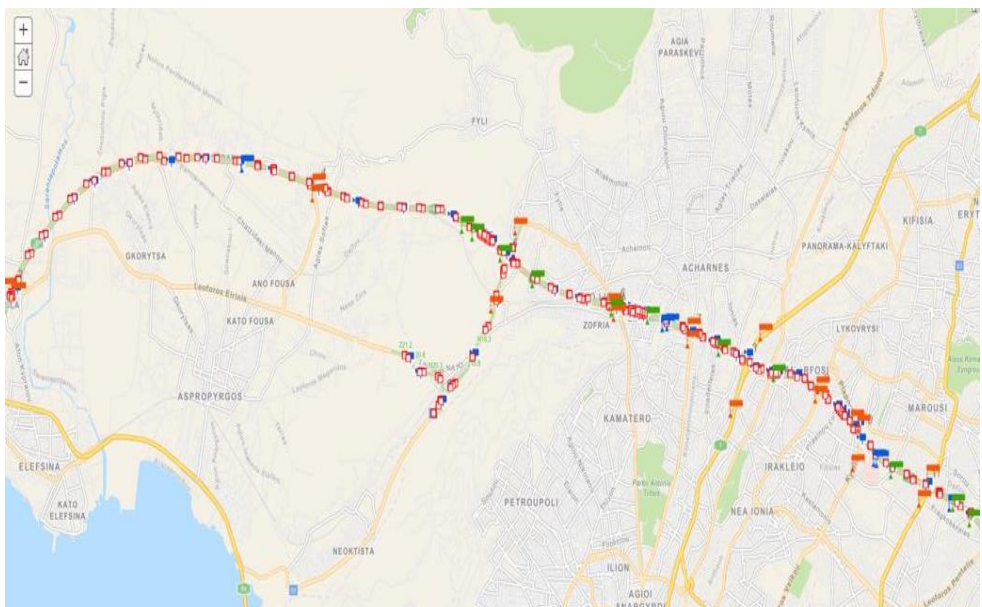

*Figure 2: Equipment installed in the Attica Tollway pilot*

The field equipment included 10 ITS-G5 Road Side Units (RSUs) along the pilot road segment - the RSUs were mainly installed in existing Variable Message Signs gantries as well as CCTV poles, so power cabling and network connectivity was readily available -, 1 mobile ITS-G5 RSU mounted on a patrol vehicle, 10 On Board Units (OBUs) operating on ITS-G5 communication protocols, and 10 OBUs operating on cellular network. All OBUs contained a Human Machine Interface (HMI) subsystem.

## 2.2 Egnatia Odos Tollway

Egnatia Odos (EO) Motorway extends to 660 km and is part of the TEN-T Core and Comprehensive Road Network Corridors. It crosses the north part of Greece from its westernmost edge (Igoumenitsa port, Ionian Sea) to its easternmost borders with Turkey (Kipoi, Evros). It's a dual carriageway motorway, with each driving direction consisting of 2 lanes (in few sections of 3 lanes) plus an emergency lane. Egnatia Odos motorway is equipped with many ITS devices and safety systems that aim at providing travel safety and comfort to the end users. Along with its 5 vertical axes of 330 km the motorway connects Greece with all its north neighboring countries (Albania, North Macedonia, Bulgaria). The section of Egnatia Odos motorway called "Kastania bypass" (**Error! Reference source not found.**) was included in the C-Roads Greece pilot. This road segment is a rural road section with continuous bridges and tunnels, which spans 26 km between two approaching intersections (IC), plus additionally 2+2km on either side of them, with Average Annual Daily Traffic (AADT) of 11,230 (HGV: 16%). It crosses a mountainous area, starting at I/C Veria at 90m altitude and ends at Polymylos I/C at 850m altitude with an average ascending gradient of +2,84%.

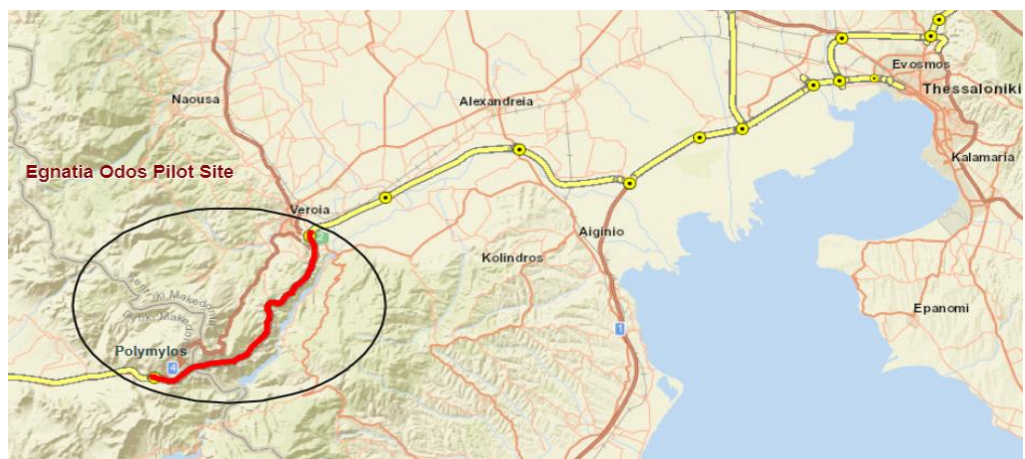

*Figure 3: Map of the Egnatia Odos Tollway pilot*

The pilot site was equipped with ITS related equipment (i.e., Variable Message Signs, Blank Out Signs (BOS), Lane Control Signs (LCS), Traffic Signals (TS), Traffic Counters with inductive loops (TC), CCTV traffic cameras, Over-height Vehicle Detectors (OHVD), Road Weather Information Systems (RWIS) & Air quality and visibility sensors in tunnels (CO/NO/VIS), etc.). This equipment was utilized as possible data gathering sources alongside the C-ITS field equipment (i.e., RSUs) that were installed during the pilot. The field equipment that was deployed on Egnatia Odos Tollway pilot for the provision of the C-ITS services included: 25 ITS-G5 RSUs, 1 mobile ITS-G5 RSU mounted on a patrol vehicle, 10 OBUs operating on ITS-G5 communication protocols, and 10 OBUs operating on cellular 4G network.

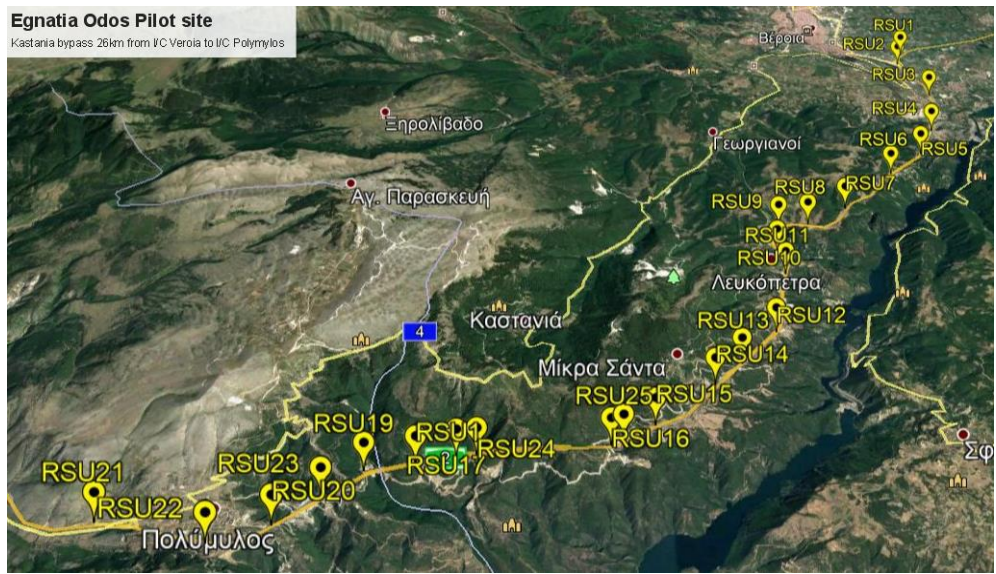

*Figure 4: Map of RSUs installations in the Egnatia Odos Tollway pilot*

## 2.3 C-ITS Services

The C-ITS services and use cases that were deployed in C-Roads Greece by the two pilots are briefly described in the following table.

*Table 1: C-ITS services of the Greek pilot*

| C-ITS Service | Description | Attica Tollway | Egnatia Odos Tollway |
|---|---|:---:|:---:|
| Hazardous Location Notification - Stationary Vehicle (HLN-SV) | Detection of a stationary/ broken-down vehicle on the road and broadcast of this information to the road users. | X | X |
| Hazardous Location Notification - Weather Condition Warning (HLN-WCW) | Provision of accurate and up-to-date information to the drivers about adverse weather conditions. | X | X |
| Road Works Warning - Lane Closure (RWW-LC) | The existence of closed lanes or lanes (including the hard shoulder) because of static roadworks and broadcast of this information to the road users. | X | X |
| In-Vehicle Signage - Embedded VMS "Free Text" (IVS-EVFT) | Provision of in-car information to the road user of type "free text". | X | X |
| In-Vehicle Signage - Shockwave damping | Sending suitably personalized messages ("alerts") to road users when the probability of a shockwave is above a threshold and when a shockwave is detected. | X | |

| | | |
|---|---|---|
| Traffic Management - Smart Routing | Provide accurate real-time information to considering the fastest route for network users. | X |
| Probe Vehicle Data - CAM Aggregation | Collect traffic data from vehicles providing information on speed, type of vehicle, weather conditions and other, that may be processed in the Traffic Control Center. | X |

## 3. Impact assessment and evaluation methodology

The methodological approach for evaluating and assessing the C-ITS services in the Greek pilot relied on the common basis for evaluation and assessment of the C-Roads Platform (Studer et al., 2020) and was adapted in accordance with the Greek pilot needs. In terms of evaluation and assessment, the core objective was to better understand the effects of providing C-ITS services. This necessitated an impact evaluation approach that could compare the observed pattern of behavior to some "counterfactual" for what would have happened without the intervention, i.e., the impacts of C-ITS services are the result of a comparison between a framework with C-ITS services that are working or activated on the equipped vehicles/ devices and other vehicles that do not have C-ITS services or have them switched off. Parameters and Key Performance Indicators (KPIs) were defined as the comparison between revealed measures with C-ITS and the baseline that was the situation without C-ITS services. The impact areas included in the evaluation and assessment approach of the Greek Pilot were user acceptance, safety, traffic efficiency, and the environment.

### 3.1 User Acceptance

Two separate questionnaires were created and distributed online to the users of the C-ITS services. Results were collected during February 2022.

The road operators' questionnaires had the purpose to evaluate whether the stuff responsible for the traffic management processes in the traffic management centers of Attica Tollway and Egnatia Odos Tollway are in favor of using the C-ITS services and more specifically the TMC Emulator, a software developed by CERTH, to assist the road operators in the operation of the C-ITS services for traffic management purposes. All participants agreed strongly with the positive expected impacts of the software on road safety increase, on the environment, and on traffic conditions improvement. All participants considered the display and exchange of information about roadworks and extreme weather conditions through the software extremely useful. Most of the participants considered the creation and exchange of C-ITS messages extremely useful, while half of the participants expected manual interactions with the software to be complex. Finally, half of the participants agreed that the software interface is user friendly.

The drivers' questionnaires had the purpose of evaluating whether drivers are in favor of using the C-ITS services that would be provided via a mobile application along the Attica Tollways and Egnatia Odos Tollway networks. Concerning RWW, most of the participants weren't familiar with the service. However, half of the participants (44,8%) considered that RWW could be a useful service for daily driving purposes and agreed strongly to daily using it, but the majority (39,7%) showed a low willingness to pay for the service. For the IVS service, almost half of the participants were familiar with the service but had never used it. They considered the service useful and stated that they would absolutely use it on a regular basis, with most of them having a neutral opinion on whether they would be willing to pay for it. Regarding HLN, most of the participants had zero knowledge about the service, but at the same time considered it useful and agreed strongly to daily using it. The highest number of the participants had a neutral opinion on willingness to pay for the service. Finally, regarding the case of Traffic Information - Smart Routing, most of the participants knew about the service but had never used it and considered it extremely useful, while most answers concerning willingness to pay were negative.

### 3.2 Transport network

The collection of real-life logs from the pilot demonstration of the services was performed via the logging mechanism which was developed by the CERTH team. When the user launched the mobile app, which provided the C-ITS services, a series of systems were initialized. In general, a connection to CERTH's broker was established, any data saved on the device from a previous session was uploaded to CERTH's SFTP server, and the location provider (Location Listener) was set up. The following table displays the real-life logged data in the two pilot locations.

*Table 2: Data collection in the Greek pilot*

| Period | Date | Attica Tollway data size - log files | Egnatia Odos Tollway data size - log files |
|---|---|---|---|
| Total data collection | 1/1/2022 - 23/8/2022 | 6.65 MB - 164 | 2.73 MB - 22 |
| Pretesting period | 1/1/2022 - 17/5/2022 | 1.56 MB - 49 | 238 KB - 6 |
| 1st baseline period | 18/5/2022 - 19/6/2022 | 134 KB - 7 | 687 KB - 5 |
| 1st treatment period | 20/6/2022 - 17/7/2022 | 1.90 MB - 56 | 1.70 MB - 8 |
| 2nd baseline period | 18/7/2022 - 15/8/2022 | 3.04 MB - 51 | 127 KB - 3 |
| 2nd treatment period | 16/8/2022 - 17/9/2022 | 13 KB - 1 | 0 |

### 3.3 Simulation experiments

The methodology for the services simulation relied on the implementation of a traffic management strategy, a Long-Term Evolution data communication network to route V2X data communications, and Cooperative Awareness Messages (CAMs) that ensure communications among C-ITS enabled vehicles by exchanging continuously data packets with information such as location, speed, etc. Moreover, the acquired information is integrated to infrastructure through a Traffic Control Server (TCS) that generates proposed messages which are taken over by the infrastructure and transmitted to vehicles via V2I communications. As CAMs indicate abrupt deceleration or sudden stop of the transmission of messages could suggest to the TCS that a crash or a hazardous incident has taken place. Finally, when the TCS detects such a conflict, it distributes the corresponding messages to the vehicles. A traffic management logic was developed for modelling the Traffic Control Server as well as the drivers' responses to warning messages using the microscopic traffic simulator SUMO. The traffic management strategy is presented in the flowchart of the following figure.

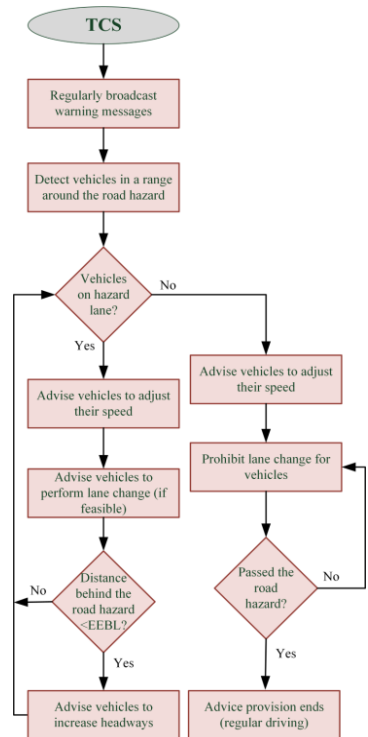

*Figure 5: Flowchart of the traffic management strategy*

The calculated KPIs included average vehicle speeds, $CO_2$ emissions, collisions, lane changes, and travel time. Regarding the number of vehicles considered in the network for simulation, two scenarios were tested for both pilot locations: 1) baseline with actual traffic demand (based on collected traffic volumes), with 500 vehicles in Egnatia Odos Tollway and 4500 vehicles (maximum capacity) in Attica Tollway, and 2) high C-ITS penetration rate with 1500 vehicles in Egnatia Odos Tollway.

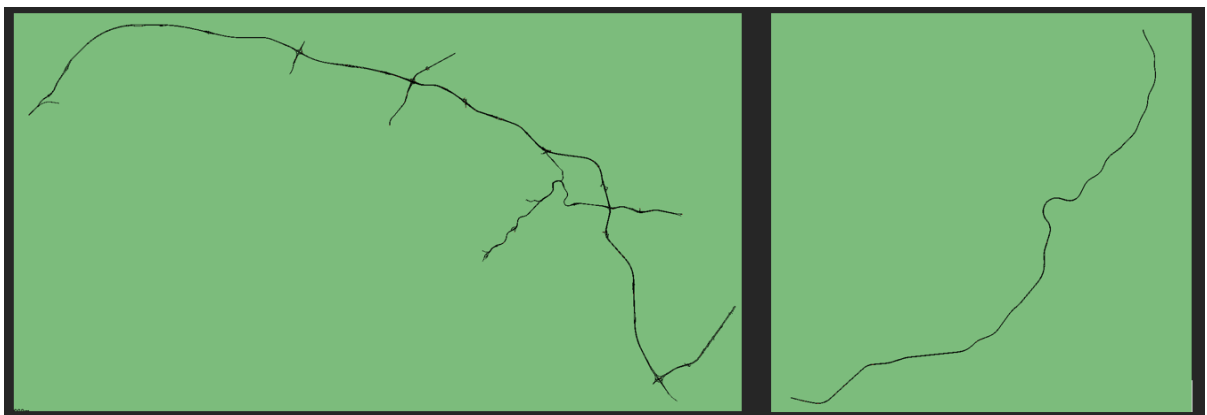

*Figure 6: Attica Tollway (left) and Egnatia Odos Tollway networks in the SUMO environment*

The C-ITS scenario describes the situation where the vehicles' speeds drop down to 0,8 500m upstream and to 0,6 when entering the edge where the event is located (speed decreases smoothly), while the manual scenario describes the situation where the vehicles' speeds drop down to 0,6 150m upstream. The number of vehicles (throughput) used in the simulation was around 4500 and is the maximum capacity in the Attica Tollway network based on actual traffic flows.

### 3.3.1 Attica Tollway network

#### 3.3.1.1 HLN-WCW (Slippery Road)

Regarding lane changes, they decreased in the C-ITS scenario, which is an expected result as in that case vehicles were informed in advance about the slippery road; hence drivers tended to avoid lane changes as it was risky. The number of collisions showed a significant decrease in the C-ITS scenario. This is an anticipated result due to the timely provision of the C-ITS messages that made drivers aware of the slippery conditions on the road and helped them drive more carefully. The HLN-WCW service could be considered that contributes to road safety at an important level. $CO_2$ emissions decreased also in the C-ITS scenario. The average vehicle speed showed a very slight increase in the C-ITS scenario; hence it could be considered that no important changes happen with the provision of C-ITS messages. Travel time remained similar as well in the C-ITS scenario, which is an anticipated result when considering that there was no change in vehicles speeds.

#### 3.3.1.2 Traffic information and smart routing (Traffic Jam Ahead Warning)

Lane changes decreased in the case of C-ITS message provision. This could be justified by the fact that vehicles in the C-ITS scenario were advised to decrease their speed in a timely manner as there was a traffic jam ahead and hence drivers didn't need to make many lane changes to avoid the event. The number of collisions decreased at a significant level in the C-ITS scenario, leading to the fact that the service could contribute to road safety. $CO_2$ emissions were decreased as well in the C-ITS scenario. This could be considered as an anticipated result due to smoother fluctuations (smoother deceleration) in vehicles speeds in the case of the C-ITS information provision. Average vehicle speed showed a very slight increase in the C-ITS scenario; hence it could be considered that no changes are observed in this indicator. The same applies for travel time. No significant changes were observed, which is something logical when considering that vehicles speeds didn't change as well.

#### 3.3.1.3 RWW-LC

The number of lane changes decreased in the C-ITS scenario. This could be justified by the fact that drivers were aware in advance about the closed lane ahead, hence they could adjust their driving behavior timely, and they didn't need to perform any sudden changes, such as many lane changes. The number of collisions was reduced significantly at the C-ITS scenario, indicating that drivers had a more attentive driving behavior due to the C-ITS messages provision, contributing this way to road safety increase. There was also a significant reduction to $CO_2$ emissions in the C-ITS scenario, leading to the conclusion that RWW-LC could have a positive impact to the environment as pollutant emissions showed a reduction due to smoother driving (smoother speed fluctuations). The indicator of average vehicle speed was lower in the C-ITS scenario. This was an anticipated result as drivers were timely informed about the lane closure and they decelerated and drove at lower speeds. Regarding travel time, this indicator showed an increase, which is something logical considering the speed decrease.

#### 3.3.1.4 HLN-OR (Slow/ Stationery Vehicle)

The indicator of lane changes didn't show a significant difference in the C-ITS scenario as it increased only slightly. This could be justified by the fact that since drivers were aware in advance of the existence of the obstacle on the road, they were able to perform lane changes timely and earlier than in the case of having no information in advance. On the other hand, the number of collisions showed a very significant decrease in the case of the C-ITS scenario indicating that the service could have a high contribution to road safety since drivers would show more attentive driving reducing this way the possibility for a collision. $CO_2$ emissions had a slight increase in the C-ITS scenario compared to the manual one. The average vehicle speed showed a decrease in the C-ITS scenario. This is since drivers were informed timely about the event, and they could adjust appropriately their driving behavior by reducing their speeds. The travel time increased significantly in the C-ITS scenario. This was due to the vehicle speed increase which was observed as well in this case.

***Table 3:*** *Attica Tollway network KPIs*

| KPI | C-ITS Service | Manual Scenario | C-ITS Scenario |
|---|---|---|---|
| Lane changes (#/km) | HLN-WCW (Slippery Road) | 0.67 | 0.58 |
| | Traffic information and smart routing (Traffic Jam Ahead Warning) | 0.67 | 0.60 |
| | RWW-LC | 0.72 | 0.67 |
| | HLN-OR | 0.84 | 0.86 |
| Collisions (#) | HLN-WCW (Slippery Road) | 14.40 | 7.65 |
| | Traffic information and smart routing (Traffic Jam Ahead Warning) | 9.70 | 5.53 |
| | RWW-LC | 16.20 | 8.70 |
| | HLN-OR | 1257.60 | 5.90 |
| $CO_2$ emissions (g/km) | HLN-WCW (Slippery Road) | 315.69 | 297.61 |
| | Traffic information and smart routing (Traffic Jam Ahead Warning) | 314.52 | 308.30 |
| | RWW-LC | 344.12 | 317.83 |
| | HLN-OR | 321.42 | 330.69 |
| Average vehicle speed (km/h) | HLN-WCW (Slippery Road) | 69.55 | 69.97 |
| | Traffic information and smart routing (Traffic Jam Ahead Warning) | 75.43 | 75.88 |
| | RWW-LC | 64.00 | 58.00 |
| | HLN-OR | 80.01 | 55.68 |
| Travel time (min/km) | HLN-WCW (Slippery Road) | 0.88 | 0.87 |
| | Traffic information and smart routing (Traffic Jam Ahead Warning) | 0.82 | 0.81 |
| | RWW-LC | 1.05 | 1.15 |
| | HLN-OR | 0.76 | 1.21 |

### 3.3.2 Egnatia Tollway network

The number of vehicles used in the simulation was around 500 for the baseline scenario, both for C-ITS and manual scenario, and around 1200 for the high C-ITS penetration scenario respectively, both for C-ITS and manual scenario.

### 3.3.2.1 HLN-WCW (Slippery Road)

Concerning lane changes performed by the vehicles, a higher number of lane changes occurred in the manual scenario for both baseline and high C-ITS penetration rate scenario, while the number of lane changes was observed to be lower in the C-ITS scenario. This could be explained because in the C-ITS scenario the vehicles are advised in the simulation to remain in the same lane and not to perform changes as all lanes are considered slippery, hence any change could be considered risky. Concerning the indicator of CO2 emissions, not a significant difference resulted in the two scenarios, but still there was a slight decrease in the C-ITS scenario for both penetration rates (500 and 1200 vehicles), leading to the conclusion that HLN - WCW does not contribute at an important level to CO2 emissions reduction. Similarly for average vehicle speed, a slight difference was observed in the two scenarios. More specifically, there was a slight increase in the C-ITS scenario for both penetration rates. This could be justified because the provision of information about the slippery road could contribute to smoother fluctuations in speed, hence speed remains higher.

### 3.3.2.2 Traffic information and smart routing (Traffic Jam Ahead Warning)

Concerning lane changes, there was an increase in the high C-ITS penetration rate scenario compared to the baseline in the case of 500 vehicles, while the opposite happens in the case of 1200 vehicles. CO2 emissions slightly increased in both scenarios, 500 and 1200 vehicles, in the case of high C-ITS penetration rate. Regarding vehicle average speed, there was quite a significant decrease in the baseline scenario (case of 500 vehicles), while a slight increase is observed in the same scenario for the case of 1200 vehicles (high C-ITS penetration rate). The same happened for travel time. Travel time was slightly increased in the baseline scenario but there is a very slight decrease in the high C-ITS penetration rate scenario.

### 3.3.2.3 RWW-LC

The number of lane changes increased in both scenarios (500 and 1200 vehicles) in the case of the C-ITS scenario. However, the increase was very low for the high C-ITS penetration rate scenario (1200 vehicles). The significant increase in lane changes in the case of the baseline scenario could be justified by the fact that in the C-ITS scenario vehicles are advised timelier about the lane closure compared to vehicles in the manual scenario, hence drivers are aware in advance about the event and try to avoid it earlier by performing more lane changes. CO2 emissions decreased in the C-ITS scenario in both cases, 500 and 1200 vehicles, but not at a significant level. Average vehicle speed showed a decrease in the C-ITS scenario in both cases (500 and 1200 vehicles). This could be due to the earlier provision of information in the case of the C-ITS messages provision where drivers start slowing down earlier and more smoothly as they are aware of the lane closure in advance. Travel time increased in the C-ITS scenario in both cases, baseline, and high C-ITS penetration rate. This is logical as the decrease in vehicle speed could lead to higher travel time.

### 3.3.2.4 HLN-OR (Slow/ Stationary Vehicle)

The number of lane changes showed a significant increase in the C-ITS scenario for 500 vehicles, but the opposite happened in the same scenario for 1200 vehicles. Lane changes were expected to increase in the C-ITS scenario as drivers were aware in advance of the event through the provision of the C-ITS messages. A significant decrease was observed in the number of collisions in the C-ITS scenarios for both cases, 500 and 1200 vehicles, hence it could be considered that HLN-OR contributes at an important level to road safety. Regarding the indicator of CO2 emissions, a decrease was observed in the C-ITS scenario in both cases. It should be mentioned that CO2 emissions decrease for HLN-OR is higher than for the abovementioned services. Average vehicle speed showed a decrease in the C-ITS scenarios, and this could be explained by the fact that since drivers are informed timely of the existence of an obstacle on the road, they are able to start decreasing their speeds earlier and more smoothly. Travel time increased in the C-ITS scenarios which is anticipated as in the same scenarios vehicles speeds show a decrease.

*Table 4: Egnatia Tollway network KPIs*

| KPI | C-ITS Service | Manual Scenario (Baseline/ High C-ITS penetration rate) | C-ITS Scenario (Baseline/ High C-ITS penetration rate) |
|---|---|---|---|
| Lane changes (#/km) | HLN-WCW (Slippery Road) | 0.43/ 0.29 | 0.36/ 0.25 |
| | Traffic information and smart routing (Traffic Jam Ahead Warning) | 0.30/ 0.21 | 0.33/ 0.20 |
| | RWW-LC | 0.30/ 0.27 | 0.40/ 0.28 |
| | HLN-OR | 0.33/ 0.31 | 0.38/ 0.27 |
| Collisions (#) | HLN-OR | 0.40/ 1.30 | 0.10/ 0.10 |
| $CO_2$ emissions (g/km) | HLN-WCW (Slippery Road) | 324.56/ 321.58 | 322.72/ 319.94 |
| | Traffic information and smart routing (Traffic Jam Ahead Warning) | 327.62/ 335.98 | 324.88/ 338.46 |
| | RWW-LC | 321.06/ 328.47 | 320/ 319.74 |
| | HLN-OR | 328.84/ 343.12 | 318.51/ 328.72 |
| Average vehicle speed (km/h) | HLN-WCW (Slippery Road) | 90.73/ 88.12 | 91.24/ 88.36 |
| | Traffic information and smart routing (Traffic Jam Ahead Warning) | 91.67/ 88.96 | 86.52/ 87.50 |
| | RWW-LC | 94.33/ 88.86 | 90.12/ 86.93 |
| | HLN-OR | 93.81/ 86.02 | 90.68/ 80.52 |
| Travel time (min/km) | HLN-WCW (Slippery Road) | | |
| | Traffic information and smart routing (Traffic Jam Ahead Warning) | 0.66/ 0.70 | 0.68/ 0.69 |
| | RWW-LC | 0.64/ 0.68 | 0.67/ 0.69 |
| | HLN-OR | 0.64/ 0.70 | 0.66/ 0.73 |

## 4. Roadmap for C-ITS large-scale deployment in Greece

The roadmap developed in the context of C-Roads Greece had the objective to provide information and resources for facilitating the large-scale deployment of Cooperative Intelligent Transport Systems (C-ITS) services in Greece. The roadmap was intended for stakeholders involved in the CCAM ecosystem, i.e., service providers, data providers, software and hardware developers and suppliers, practitioners responsible for road network management and operations (public authorities, road authorities and operators, and traffic management centers). The roadmap provides insights into the C-ITS ecosystem by focusing on three different perspectives that capture core stakeholder types: 1) service providers, 2) end users, and 3) governance. The context of the roadmap expresses a single vision of the C-ITS deployment in Greece within a time horizon of 2030. The structure of the roadmap is as follows:
Core themes: Services, Infrastructure, Vehicles, and Society as a whole.
Individual topics under each core theme.

## 4.1 Services

### 4.1.1 Personal and inclusive mobility

The integration of people's needs in C-ITS technologies should consider all needs and requirements resulting from VRUs and more importantly from peoples with disabilities, to define new services that capture future mobility needs. Based on that, a milestone to be achieved by 2028 is the offering of commercial CCAM solutions for all travelers, followed by the provision of sustainable services in wider areas (urban, peri-urban, and rural) and not only large cities.

### 4.1.2 Freight transport and logistics

The first step is the adoption and continuation of small-scale tests and pilots in the context of research programs and other initiatives, that will foster investment opportunities for testing facilities. The goal for 2026 is the wide integration of connected and automated technologies in freight transport modes, which will be followed by the establishment of licensing frameworks for last-mile deliveries, hence allocating curb space for CCAM-enabled vehicles. Finally, the establishment of a national freight traffic control and management service by 2030 will facilitate the coordination and consolidation of freight and personal transport.

### 4.1.3 Multimodal mobility

The integration of CCAM technologies in multimodal transport summarizes the two abovementioned topics highlighting first the need for pilot trials that will include both C-ITS services for individual travelers and commercial fleets. By 2028 commercial CCAM solutions for multimodal transport should be widely accessible leading gradually to scheduling of integrated journeys within the context of combined C-ITS services and Mobility-as-a-Service (MaaS) solutions.

## 4.2 Infrastructure

### 4.2.1 Testing

The development of a Testbed in Greece could serve as a national center for innovation and research in CCAM technologies, including settings and facilities in a wide range of highways, urban, and peri-urban networks, as well as controlled and public road segments where testing activities could be performed both in physical and virtual environments. Adding digital capabilities to the Testbed could be accompanied by activities for the development and establishment of a national Public Key Infrastructure (PKI) by 2029 that will set the basis for the development of a center of excellence in the domain of cybersecurity for CCAM technologies.

### 4.2.2 Intelligent Traffic Management

The identification of new travel patterns generated from the use of CCAM services constitutes an integral part of defining new planning methods and metrics, to increase network efficiency through the implementation of innovative operational models. The development of a national connected vehicles services platform will serve as preliminary action for the establishment of a nationally interconnected traffic management system which will orchestrate, monitor, and control multimodal CCAM and traditional services.

### 4.2.3 Road infrastructure

A significant element in defining the evolution of road infrastructure towards the wide adoption of C-ITS technologies is the identification of new business models. Alongside, safety requirements should be clearly defined and new standards for infrastructure design and safety services for CCAM-enabled vehicles should be developed. From an assets' perspective, the concept of a digital highway where all roadside equipment will be entirely digitized, could start with the recognition of requirements and specifications for digital road infrastructure which will be put in place by 2029 through the introduction of digitally signalized road networks. Digital traffic rules defined by 2030 will specify how various CCAM services should be prioritized on the road network.

### 4.2.4 Digital infrastructure

An important first step from 2023 is enabling cross-sector data sharing which is comprised by the definition of sharing agreements between vehicles and the infrastructure, data processing and storing. These will form new standards for open data and governance rules. One step further will be the establishment of cyber secure data services in the context of which data will be securely and safely handled for optimum services provision. As digital infrastructures extend and matures investments into national digital twins can initiate from 2028 to serve as environments for routing optimization, assets management, real-time mapping information visualization, and various traffic control layers display. Finally, wide access to commercial data hubs for any stakeholder in the CCAM ecosystem should be established by 2030.

### 4.2.5 Communication networks

The establishment of a nationwide network communications coverage plan will include communication solutions for critical elements of vehicles' operations, thorough business cases for emerging communications models, and specifications on the required coverage and network resilience for robust infrastructural solutions. As communications coverage roll-out progresses, wider areas of the road network will be equipped.

## 4.3 Vehicles

### 4.3.1 Sensors' equipment

Considering the most optimistic scenario, where CCAM services and self-driving vehicles are fully integrated in Greek road networks by 2030, sensors equipment development constitutes one of the most important elements towards this direction. An objective set for 2025 is the provision of low cost and high precision and accuracy components which can be either imported in the Greek market or designed and developed from local industry. Next steps from 2025 include the adoption and establishment of nationally agreed common standardization processes and calibration methods, which will allocate efficient and harmonized collection and processing of reliable and accurate data. In parallel the research domain in recognition testing can be further extended and enhanced with deeper and more sophisticated initiatives generating outcomes related to validation processes. Finally, the target for 2030 is the development of advanced sensors that will support higher levels of automation.

### 4.3.2 Driver behavior

The new requirements focus mainly on safety needs that will ensure that ergonomics and design features capture them. Further research on the domain of HMI should be supported with the purpose of identifying common standards and specific HMI features and design that should be integrated to CCAM-enabled vehicles.

### 4.3.3 Connected cars

From 2023 Greece should strongly participate in standardization activities, to be able to conform with pan European or even global standards that will be harmonized and enable the implementation of the latest and most innovative Vehicle-to-Everything (V2X) services. Greater amounts of data could be exploited from digital twins that will produce new information leveraged by 2030 for integrated journeys scheduling.

### 4.3.4 Automated cars

Starting from now Greece should invest in research activities that will focus on strengthening the ADSs' automation capabilities and contribute to the standardization of Operational Design Domains (ODDs). In 2024 the first commercial automation elements could start to integrate in freight and passenger vehicles. Such elements could be enhanced with more advanced features and the new functionalities could begin from 2026 to integrated in the vehicles with reliable and valid update procedures. Another important factor affecting this core theme, is Artificial Intelligence (AI) and Machine Learning (ML), hence certified tools, algorithms, and techniques should be defined and standardized from 2028 onwards, allocating by 2030 the integration of full automation features in all transport modes.

## 4.4 Society as a whole

### 4.4.1 Training and skills

The first step towards this goal is the understanding and the existing knowledge on CCAM technologies and the identification of gaps that should be filled, to develop strong skill sets. For this purpose, national training activities, initiatives, and campaigns should be intensified by 2025. These initiatives could serve as the starting point for the establishment of a national CCAM center of excellence by 2029 that will support courses, training programs, and research projects targeting to equip all stakeholders participating in the CCAM ecosystem.

### 4.4.2 Market ecosystem

The establishment of a funding and investments ecosystem from 2023 will promote the active engagement of government, venture capitals, and private equity. Moreover, investments in small and medium-sized enterprises (SME) will strengthen the conditions for local growth. Complementary to that, new approaches to mobility and infrastructure business models will set the ground for a mature and collaborative market ecosystem capable of entering global CCAM markets by 2030.

### 4.4.3 Public acceptance

The development of a concrete education plan from 2024 could ensure that such initiatives will be executed in a well-programmed and coordinated manner, building consistent knowledge for the new technologies. At the same time and with the introduction of commercial CCAM solutions to the market, an increased exposure to these services could be achieved from 2026 onwards, leading gradually to discard any stereotypes and misconceptions around CCAM technologies and enabling wide public acceptance by 2030.

### 4.4.4 Legislation and insurance framework for CCAM

The establishment of efficient safety regulations and insurance models relies significantly on data, hence appropriate data sharing should be ensured. Based on that the necessary legislative adjustments should be made by mid-2026 and will be followed by updates of the regulatory and legal framework for insurance. This mature and robust framework will enable and promote the provision of lower cost insurance packets for CCAM-enabled vehicles by 2030.

### 4.4.5 Licensing and use

The first step for establishing the necessary framework for licensing and use of such vehicles is the alignment with global standards and the definition of the necessary updates on the legislative and regulatory context. Starting within 2026 safety protocols for CCAM services should be defined, to ensure that the actual use of the connected and automated vehicles will take place in a secure and safe environment. The final step to be achieved by 2030 is the establishment of a national licensing scheme for CCAM services.

### 4.4.6 Vehicles trials

An initial framework for CCAM trials approval could start to develop as of now, defining the processes under which vehicles will be tested to meet minimum safety operation and functionalities standards. This initial framework could be further enhanced and lead to a mature national trials approval scheme which could be integrated by 2030 into harmonized international approval processes.

## 5. Conclusions

The evaluation and impact assessment results were in general positive concerning both user acceptance and the impact of the services at transport level. Regarding the expectations of the users, the majority showed that they are in favor of using the C-ITS services and they expected that many aspects related to driving and transport would be improved by using the services. The only factor that remained low was the willingness to pay for the services as most of the users had a neutral opinion about it or were negative to such an option. Concerning the impact of the services at transport level, the results generated by the simulation experiments indicated that most of the services can have a positive contribution to road safety, as they lead to collisions reduction and more careful lane changes. Also, a positive impact could be expected with regards to traffic efficiency, since C-ITS messages provision led to smoother speed fluctuations. Finally, the use of the services contributed also in some use cases to $CO_2$ emissions reduction, indicating the potential of C-ITS to lead to negative environmental impacts reduction. To generate benefits of C-ITS technologies for all types of stakeholders, it becomes apparent that the development of roadmaps beyond initial deployments is required. The scope of the roadmap developed within C-Roads Greece is to provide direction for the future of C-ITS and CCAM in Greece. The intention of the roadmap is also to enable multiple sectors to establish new relationships and achieve cross-sector and cross-industry collaborations. The context of the roadmap expresses a single vision of the C-Roads Greece project consortium towards activities and milestones necessary for the large-scale C-ITS deployment in Greece within a time horizon of 2030.

## 7. References-Bibliography